\documentclass[doublecol]{epl2} % for 2 columns style without line numbers
\usepackage{graphicx}
\usepackage{color}
\usepackage{epsfig}
\usepackage{amssymb}
\usepackage{amsfonts}
\usepackage{amsmath}

\title{Measurement of the ionization yield of nuclear recoils in liquid
argon at 80 and 233 keV}
\author{A. Bondar\inst{1,2} \and A. Buzulutskov\inst{1,2}\footnote{Corresponding author. Email:
A.F.Buzulutskov@inp.nsk.su} \and A. Dolgov\inst{2} \and E.
Grishnyaev\inst{1} \and S. Polosatkin\inst{1,2} \and L.
Shekhtman\inst{1,2} \and E. Shemyakina\inst{1,2} \and A.
Sokolov\inst{1.2}} \shortauthor{A. Bondar \etal}

\institute{
  \inst{1} Budker Institute of Nuclear Physics SB RAS, Novosibirsk, 630090, Russia\\
  \inst{2} Novosibirsk State University, Novosibirsk, 630090, Russia
}


\abstract{The energy calibration of nuclear recoil detectors is of
primary importance to rare-event experiments such as those of direct
dark matter search and coherent neutrino-nucleus scattering. In
particular, such a calibration is performed by measuring the
ionization yield of nuclear recoils in liquid Ar and Xe detection
media, using neutron elastic scattering off nuclei. In the present
work, the ionization yield for nuclear recoils in liquid Ar has for
the first time been measured in the higher energy range, at 80 and
233 keV, using a two-phase Cryogenic Avalanche Detector (CRAD) and
DD neutron generator. The ionization yield in liquid Ar at an
electric field of 2.3 kV/cm amounted to 7.8$\pm$1.1 and 9.7$\pm$1.3
e$^-$/keV at 80 and 233 keV respectively. The Jaffe model for
nuclear recoil-induced ionization, in contrast to that Thomas-Imel,
can probably consistently describe the energy dependence of the
ionization yield.}

\begin{document}

\maketitle

\section{Introduction}

The energy calibration of nuclear recoil detectors is of
primary importance to rare-event experiments
\cite{NobleRev} such as those of direct dark matter
search
\cite{Xenon10,Xenon100,Zeplin3,Warp,Lux,ArDM,Darkside,Dama,Cogent,Crest,Cdms}
and coherent neutrino-nucleus scattering \cite{CoNu1,CoNu2}. Such a
calibration, in particular in liquid Ar and Xe detection media, is
usually performed by measuring the ionization yield and
scintillation efficiency of nuclear recoils, using neutron elastic
scattering off nuclei (the latter imitating the interaction with
dark matter particle or coherently scattered neutrino).
While for liquid Xe there is an ample of experimental data on such
yields \cite{LXeYield1,LXeYield2,LXeYield3}, little is known  about
the ionization yield \cite{Joshi,Cao} and scintillation efficiency
\cite{LArScint} in liquid Ar.

Recently the first results on the ionization yield of nuclear
recoils in liquid Ar have been presented, in the lower energy range:
at 6.7 keV \cite{Joshi} and 17-57 keV \cite{Cao}. In the present
work, the ionization yield of nuclear recoils in liquid Ar has for
the first time been measured at higher energies, at 80 and 233 keV.
These results complement those measurements conducted at
lower energies and thus might be relevant to the future dark matter
search experiments \cite{ArDM,Darkside} and to thorough
understanding of the ionization yield in liquid Ar. The
measurements were performed using a dedicated two-phase Cryogenic
Avalanche Detector (CRAD) \cite{CRADRev} and DD neutron generator.
This study was conducted in the course of the development of
two-phase CRADs of ultimate sensitivity for rare-event experiments
\cite{CRADRev}, based on thick Gas Electron Multiplier (THGEM)
technique \cite{THGEMRev}, with charge \cite{CRAD-THGEM} and
combined charge/optical
\cite{NIRStudy1,NIRStudy2-CRADProject,CRAD-GAPDMatrix} readout.

The ionization yield measured in experiment is defined as follows:

\begin{equation}
\label{eq.1} Q_y=n_e/E_0
\end{equation}
Here  $n_e$ is the number of electrons escaping recombination with
positive ions; it depends on the energy deposited by a recoil
nuclear in the liquid ($E_0$) and on the electric field in the
liquid ($\mathcal{E}$). $n_e$  is always smaller than the initial
number of ion pairs produced in the liquid by a nuclear recoil
($N_i$). In the absence of a complete recombination model, it is
generally accepted that the following parametrization
works well \cite{NobleRev,Barabash93}:

\begin{equation}
\label{eq.2} n_e=\frac{N_i}{1 + k/\mathcal{E}}
\end{equation}
where $k$ is a fitting constant.

Equations 1 and 2 are valid for both electron recoils,
induced by electron or gamma-ray irradiation, and nuclear recoils;
it is conventional to refer to the corresponding recoil
energy in units of keVee (electron-equivalent) and keVnr. The goal
of the present study is to measure $Q_y$ for nuclear recoils in
liquid Ar at certain energies and electric field and to compare it
with that calculated using existing theoretical models.

\section{Experimental setup}
The experimental setup was similar to that used in our previous
studies of two-phase CRADs in Ar with THGEM-based charge readout
\cite{CRAD-THGEM,NCount}. In the two-phase CRAD with charge readout,
the primary ionization (electron) charge produced in the liquid by a
particle, drifts towards a liquid-gas interface under an electric
field, where it is emitted into the gas phase and further multiplied
with a double-THGEM multiplier, in a proportional
(electron-avalanching) mode. The experimental setup included a
vacuum-insulated cryostat with a 9 l volume cryogenic chamber. The
chamber consisted of a cathode mesh, immersed in a 0.8 cm thick
liquid Ar layer, and a double-THGEM assembly with an active area of
10$\times$10 cm$^2$, placed in the gas phase above the liquid. The
detector was operated in a two-phase mode in the equilibrium state,
at a saturated vapour pressure of 1.0 atm and at a temperature of 87
K. In this study, the charge gain of the double-THGEM multiplier was
equal to 250; the electric field within the liquid was equal to
$\mathcal{E}=2.3$ kV/cm. The Ar was purified by an Oxisorb filter,
providing an electron life-time of $\ge$13 $\mu$s in the liquid,
corresponding to the electron drift path before attachment of $\ge$5
cm.

The primary ionization charge in liquid Ar was produced by either
X-rays from a $^{241}$Am source having among others a 59.5 keV line,
or DD neutron generator. The detector was irradiated from outside,
practically uniformly across the active area, through two Al windows
located at the chamber's bottom.

To produce neutrons a specially designed  neutron generator was used
that continuously emitted monoenergetic neutrons (not collimated)
with the kinetic energy of 2.45 MeV obtained in the DD fusion
reaction \cite{NScatProp, NGen}. The neutron flux was of the order
of 10$^4$ s$^{-1}$ over full solid angle.  The neutron generator
(operated at 80 kV and wrapped in a 1 cm thick Pb screen to suppress
bremsstrahlung gamma-rays) was placed underneath the two-phase CRAD
at a distance of about 10 cm from the active volume.

The charge signals from the two-phase CRAD were recorded from the
last electrode of the second THGEM using charge-sensitive
amplifiers, with an overall time constant of 10 $\mu$s. The signals
were digitized and memorized for further off-line analysis with a
LeCroy~HRO~66Z digital oscilloscope.

The trigger was provided by the signals from the two-phase CRAD
itself at a detection threshold equivalent to the energy of 8 keVee,
well above the electronic noises. At this threshold the typical
detector rate under neutron irradiation was of the order of 10
s$^{-1}$, well below the DAQ saturation rate limit. Each
experimental session was composed of several measurement runs: that
with neutron generator on, that with neutron generator off (to
measure the cosmic and detector internal backgrounds) and those of
the charge-scale and energy-resolution calibration using a 59.9
X-ray line from $^{241}$Am source.

In addition, a neutron scintillation counter made of
stilbene (C$_{14}$H$_{12}$) was enabled \cite{NCount}; it was placed
close to the CRAD active volume, just underneath the neutron
generator. The counter could effectively separate neutrons from
gammas using a pulse-shape analysis and thus estimate the gamma-ray
background due to ($n,\gamma$) reactions in the two-phase CRAD.

\section{Experimental results}
\begin{figure}
\includegraphics[width=0.9\columnwidth]{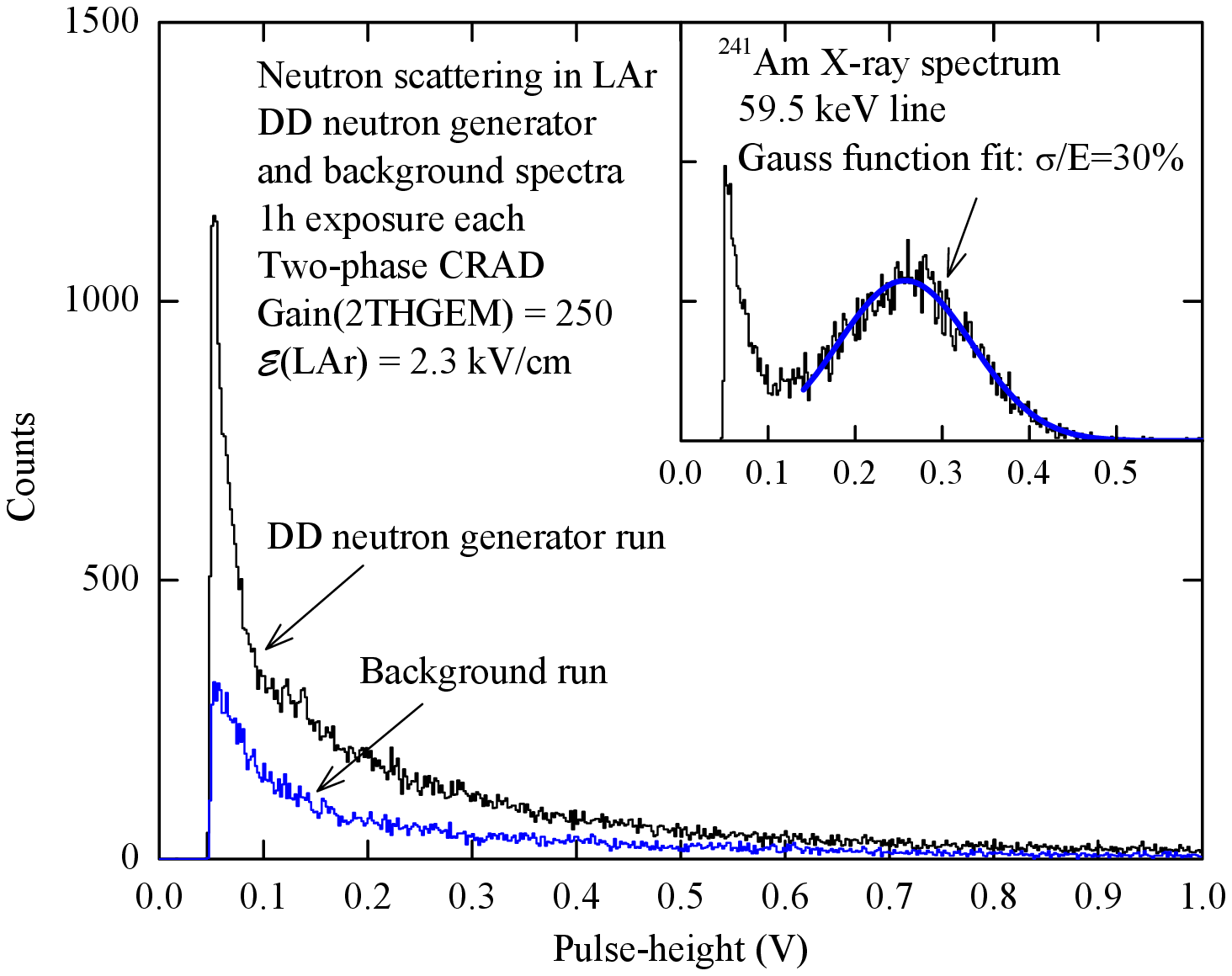}
\caption{Pulse-height distributions of the two-phase CRAD signals in
the measurement runs when the neutron generator was on (DD neutron
generator run) and off (Background run). The run time exposure was 1
h. In the inset, the pulse-height distribution induced by X-rays
from $^{241}$Am source, in one of the calibration runs, is shown.}
\label{SpectrRaw}
\end{figure}

Fig.~\ref{SpectrRaw} shows the raw amplitude spectra in the
measurement runs of the most successful session: two pulse-height
distributions, obtained in the runs when the neutron generator was
on and off, are presented. To obtain the real neutron
scattering-induced spectrum, the latter should be subtracted from
the former. To calibrate the amplitude scale in terms of the primary
ionization charge, the detector was also irradiated with X-rays from
$^{241}$Am source in the calibration runs: see the inset in the
figure. In addition in these calibration runs the amplitude
resolution of the detector was measured using the 59.5
keV line: it amounted to $\sigma$/E=30\%. The resolution
is expected to be reasonably spatially uniform and
practically independent of the energy, being mostly defined by the
gain non-uniformity over the THGEM area and the fluctuations of the
slow component of an electron emission though the liquid-gas
interface \cite{TPAr-Emission}.

In Eq.~1 one should first determine $n_e$ from the spectra of
Fig.~\ref{SpectrRaw}. For this the pulse-height amplitude in the
figure should be normalized to that of 59.5 keV peak and then
converted to the primary ionization charge. For the latter one
should know, in turn, the ionization yield for 59.5 keV X-rays in
liquid Ar. Since the experimental data on this quantity were not
available in the literature, we conducted a special intermediate
study on the ionization yields for electron-equivalent recoils,
namely for X-rays and electrons absorbed in liquid Ar and Xe.

\begin{figure}
\includegraphics[width=0.9\columnwidth]{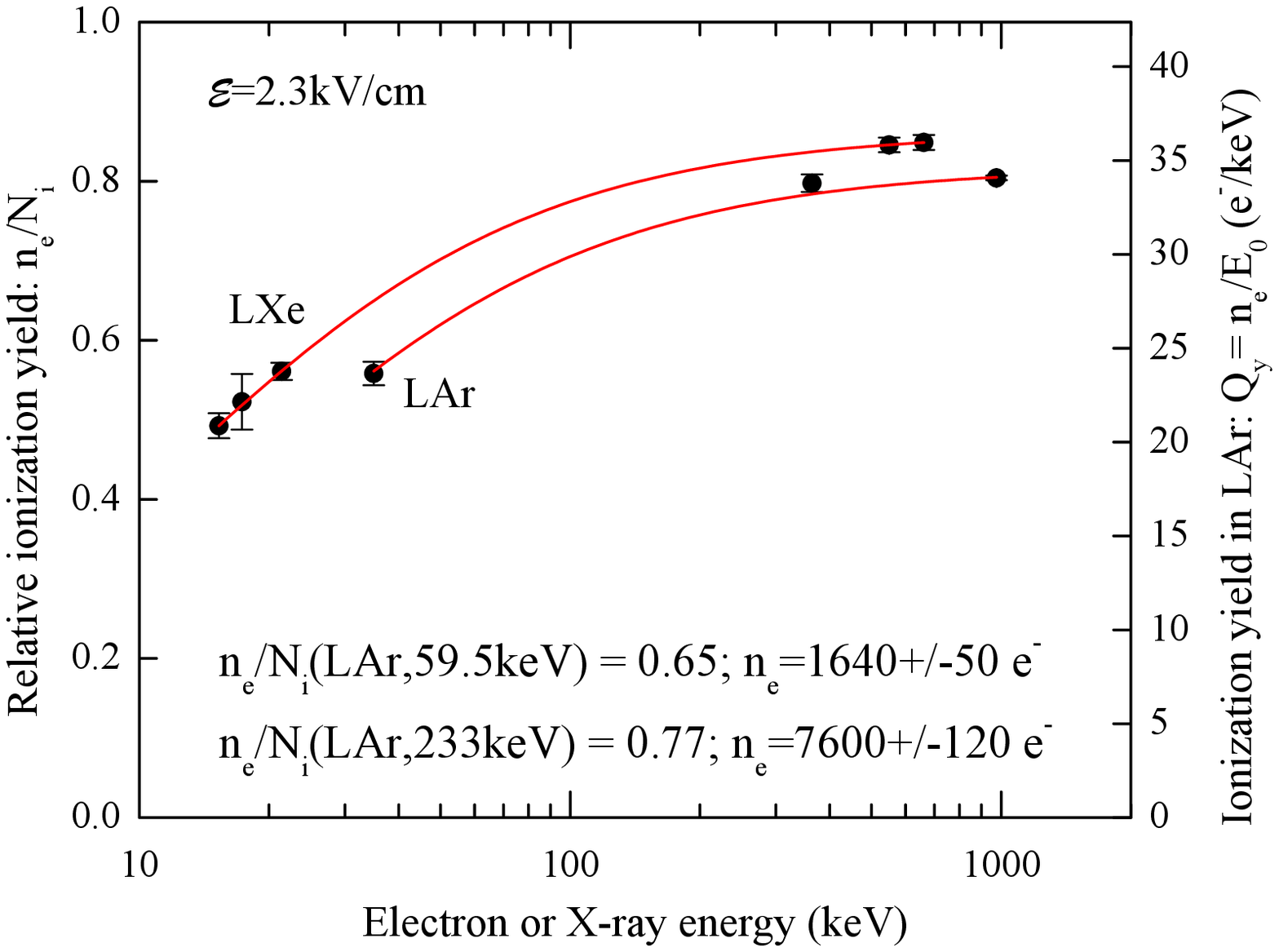}
\caption{Relative (left scale) and absolute (right scale) ionization
yields in liquid Ar and Xe for electron-equivalent recoils, namely
for X-rays and electrons, as a function of the X-ray or electron
energy, at an electric field of 2.3 kV/cm. The experimental data
points in liquid Xe were obtained from refs. \cite{Voronova89}
(15.3, 17.3, and 21.4 keV), \cite{Shibamura75} (550 keV) and
\cite{Obodovskiy79} (662 keV); in liquid Ar - from the present work
(35 keV), from refs. \cite{Scallettar82} (364 keV) and
\cite{Shibamura75} (976 keV). The data on the constant $k$ of Eq. 2
were used from table 2.6 of ref. \cite{Barabash93}, if necessary, to
calculate the relative ionization yield at a field of 2.3 kV/cm for
those experimental data points.} \label{EIonYield}
\end{figure}

The results are presented in Fig.~\ref{EIonYield}, showing the
relative ($n_e/N_i$) and absolute ($Q_y=n_e/E_0$) ionization yields
in liquid Ar and Xe as a function of energy. In liquid Xe, where
there are enough experimental data in the energy range of interest,
the relative ionization yield dependence on energy is perfectly
described by a function $n_e/N_i=1/(1+a/E_0)$ with a single
parameter $a$. Accordingly, in liquid Ar we used the similar
function, firstly, to describe the experimental data and, secondly,
to extrapolate the data to the energy point of interest. In liquid
Ar however there was a lack of data in the low energy domain. In
this domain the data point at 35 keV was obtained ourselves in a
dedicated experiment using a small liquid Ar TPC and pulsed X-ray
tube, previously used in the study of NIR scintillations in liquid
Ar (see \cite{NIRStudy1} and Fig.~7 in ref.
\cite{NIRStudy2-CRADProject}); here the X-ray energy was accurately
calculated using a dedicated computer program for the
given X-ray tube and TPC geometry.

Finally, to calculate $n_e$ (as well as $Q_y$) from $n_e/N_i$, we
used a $W$-value (energy needed to produce one ion pair) and its
definition, $W=E_0/N_i$: $W=23.6$ eV in liquid Ar (see table 2 in
ref. \cite{NobleRev}). The results of this intermediate study are
presented in Fig.~\ref{EIonYield}: in particular, for 59.5 keV
X-rays the number of detected electrons in liquid Ar at a field of
2.3 kV is predicted to be $n_e=1640\pm50 e^-$.

\begin{figure}
\includegraphics[width=0.99\columnwidth]{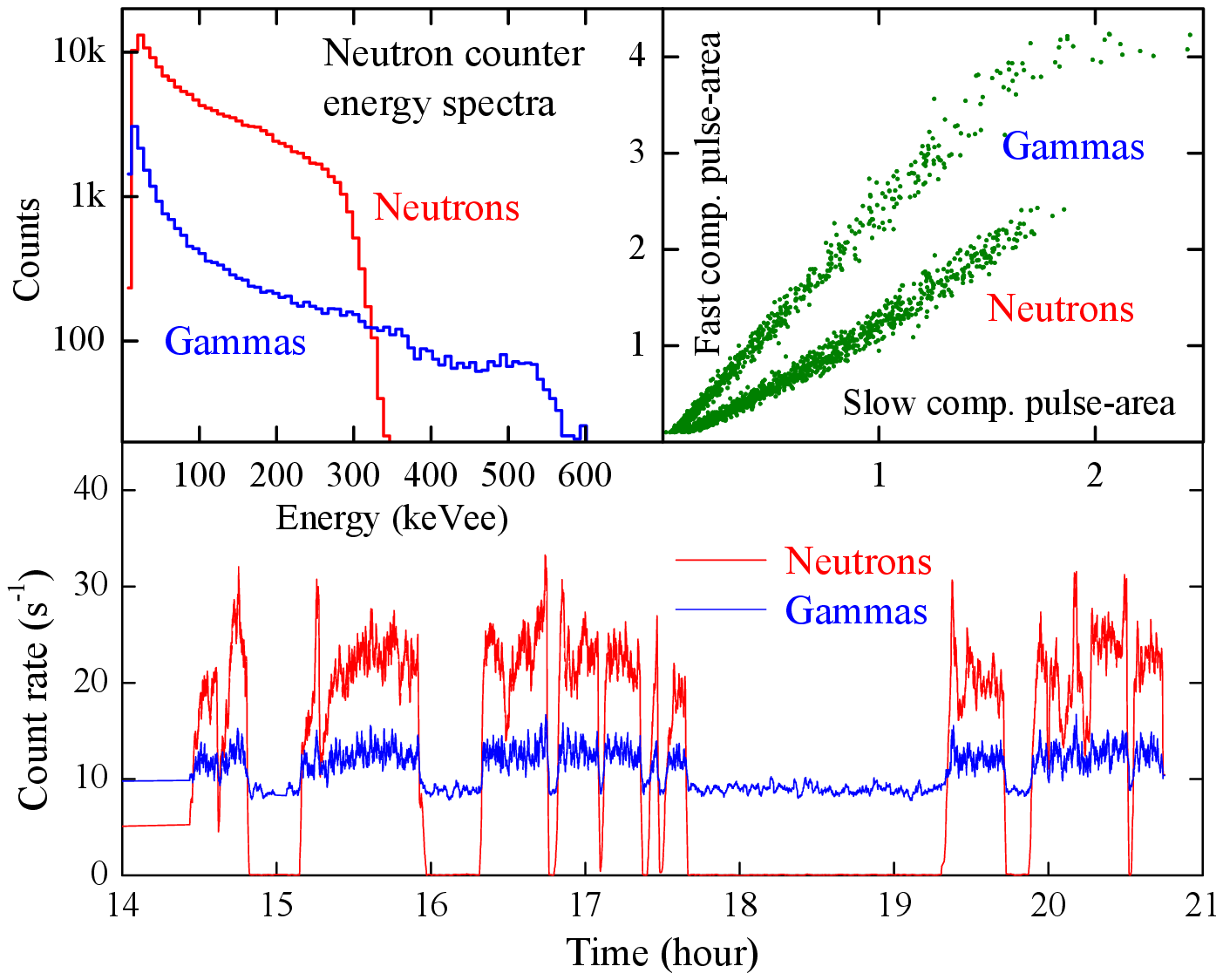}
\caption{Data from the neutron counter. Top-right panel:
neutron/gamma-ray separation using the fast and slow pulse
components of the stilbene-scintillator signal. Bottom panel: the
neutron and gamma-ray count rate evolution during the experimental
session with several measurement runs when the neutron generator was
either on or off. Top-left panel: the energy distributions of the
neutron- and gamma-ray-induced events, the contribution of the
latter being equal to 15\% of that of the former. Here
the energy scale calibration was performed using the neutron
spectrum end-point, corresponding to proton recoils with the energy
of 2.45 MeV, and the scintillation quench factor in stilbene for
protons of 0.13 \cite{QFStilbene}.} \label{NCounter}
\end{figure}

After subtracting the background-run contribution, the amplitude
distribution still contains the gamma-ray background, resulted from
($n,\gamma$) reactions in surrounding materials. This contribution
was experimentally determined using the data from the neutron
counter: see Fig.~\ref{NCounter}. Firstly, the neutron signals were
effectively separated from those of gamma-rays applying the
pulse-shape analysis (see top-right panel). Then the gamma-ray
contribution was determined comparing the neutron and gamma-ray
count rates during the experimental session, namely when the neutron
generator was on and off (see the bottom panel): it was equal to
15\% of the neutron count rate. Finally, the energy spectra for
neutrons and gamma-rays were measured in the counter (see the
top-left panel). The gamma-ray spectrum here is defined by electron
recoils in stilbene from the carbon atoms due to Compton scattering
of gamma-rays produced by surrounding H, C, Ar, Fe, Pb, etc nuclei
with the energies above 0.6 MeV.

\begin{figure}
\includegraphics[width=0.9\columnwidth]{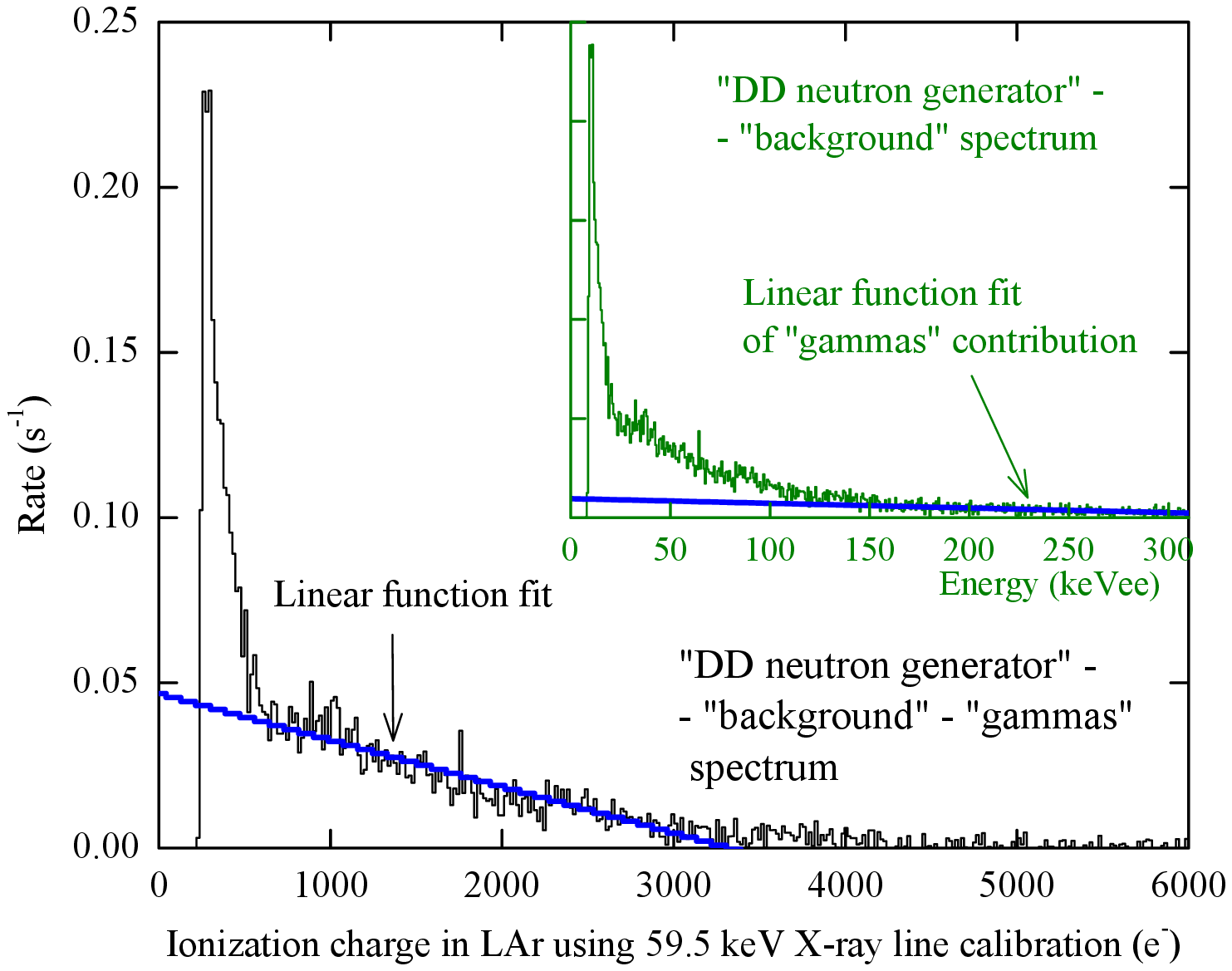}
\caption{Ionization charge distribution in liquid Ar induced by
nuclear recoils due to neutron scattering, at an electric field of
2.3 kV/cm, obtained from Fig.~\ref{SpectrRaw} using 59.5 keV X-ray
line calibration and subtracting background-run and gamma-ray
contributions. In the inset, the energy distribution is shown after
subtracting the background-run contribution, but before subtracting
the gamma-ray contribution, the latter being fitted by a linear
decreasing function. Here the energy scale calibration was performed
using the 59.5 keV X-ray line.} \label{SpectrNeutron}
\end{figure}

At these photon energies the spectrum of electron recoils
due to Compton scattering at the given recoil energies can be
approximated by a linear decreasing function with energy (see Fig.
10.1 in ref. \cite{Knoll}). Accordingly in liquid Ar the gamma-ray
contribution due to Compton scattering appeared in the energy
spectrum as a long tail extending to higher energies, fitted by a
linear decreasing function: see the inset in
Fig.~\ref{SpectrNeutron}. The contribution of gamma-ray background
determined that way amounted to about 45\% of that of neutrons. This
is close to that expected from the neutron counter data: the Compton
effect in Ar is enhanced in proportion to the atomic number when
compared to that in C, i.e. roughly by a factor of 3, while the
neutron-nucleus cross-sections almost do not change.

Fig.~\ref{SpectrNeutron} shows the desired ionization charge
spectrum in liquid Ar induced by nuclear recoils due to neutron
scattering; the spectrum is obtained from Fig.~\ref{SpectrRaw} using
59.5 keV X-ray line calibration and subtracting the background-run
and gamma-ray contributions. The next step is to compare this
experimental spectrum to that of theoretical.

\begin{figure}
\includegraphics[width=0.9\columnwidth]{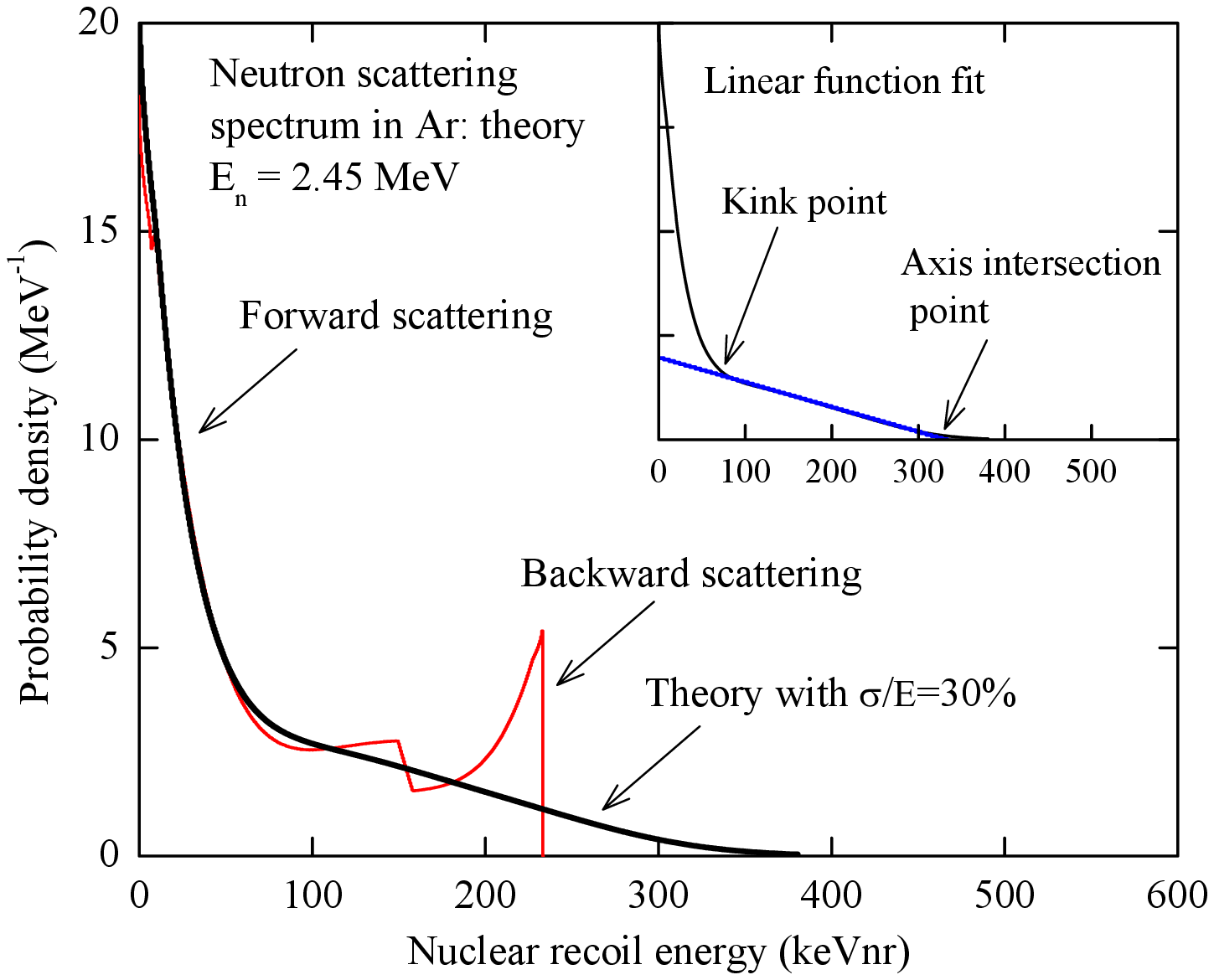}
\caption{Nuclear recoil energy distribution induced by scattering of
neutrons with the energy of 2.45 MeV off $^{40}$Ar nuclei, computed
theoretically (red line). The convolution of the theoretical
spectrum with the energy resolution of the two-phase CRAD obtained
in experiment (30\,\%), is also shown (black line). In the inset,
two characteristic points of the spectrum with respect to the linear
function fit of the backward scattering component are indicated:
that of the axis intersection at the spectrum end-point and that of
the spectrum kink at the transition between the forward and backward
scattering components.} \label{SpectrTheory}
\end{figure}

The theoretical  spectrum was computed using simulation code {\it
Scattronix} developed in the Budker INP~\cite{NSpectr} and
differential cross-sections of elastic and inelastic neutron
scattering \cite{NData}. For the $^{40}$Ar nuclei irradiated by
neutrons with the energy of 2.45 MeV it is shown in
Fig.~\ref{SpectrTheory}. One can see that the maximum recoil energy
for the Ar nuclei, achieved for backward neutron scattering at
180$^o$, is 233 keV. The convolution of the theoretical spectrum
with the energy resolution of the two-phase CRAD obtained in
experiment (30\,\%), is also shown; just this spectrum should be
compared to that of experimental. Two components of neutron
scattering are distinctly seen in the figure: those of the forward
and backward scattering, respectively represented in the
distribution by the peak at lower energies and the shoulder at
higher energies. These two components are well reproduced in the
experimental spectrum in Fig.~\ref{SpectrNeutron}. It is interesting
that the backward scattering component (the shoulder) in both the
experimental and theoretical spectra is well approximated by a
linear function: see Fig.~\ref{SpectrNeutron} and the inset in
Fig.~\ref{SpectrTheory}.

One cannot directly fit the theoretical spectrum to that of
experimental in order to obtain the ionization yield, due to the
unknown energy dependence of the yield. On the other hand, one can
select the characteristic points at certain nuclear recoil energies
in both the experimental and theoretical spectra and calculate the
ionization yield at these points according to Eq.~\ref{eq.1}. Namely
at these points, one should divide the ionization charge value (in
the experimental spectrum) by that of the nuclear recoil energy (in
the theoretical spectrum). Fortunately, the mathematically justified
choice of two such points can be done with respect to the linear
function fit of the backward scattering component (see the inset in
Fig.~\ref{SpectrTheory}). The first point, characterizing the
spectrum end-point, is that of the horizontal axis intersection of
the linear function. The second point, characterizing the transition
between the forward and backward scattering components, is that of
the spectrum kink, defined as the spectrum deflection from the
linear function fit of the backward component. The nuclear recoil
energies which should be attributed to these two points are those of
233 keV and 80$\pm$2 keV, respectively.

Consequently, the ionization yield values at these two energies
measured in that way are as follows: $Q_y=9.7\pm1.3\,e^-$/keV at 233
keV and $Q_y=7.8\pm1.1\,e^-$/keV at 80 keV. These are presented in
Table 1, showing also the appropriate statistic and systematic
uncertainties and their quadratic sum (total error).

\begin{table}
\caption{Ionization yields ($Q_y$) of nuclear recoils in liquid Ar
with uncertainties measured at an electric field of 2.3 kV/cm}
\label{tab.1}
\begin{center}
\begin{tabular}{lcccc}
Energy & $Q_y$ & Statistical & Systematic & Total\\
(keV)& ($e^-$/keV) & & & error \\ [2mm]\hline
233  & 9.7 & 0.29 & 1.3 & 1.3\\
80$\pm$2   & 7.8 & 0.24 &  1.05 & 1.1

\end{tabular}
\end{center}
\end{table}

In addition to the ionization yield, the ionization efficiency,
sometimes called the ionization quench factor, can be determined.
The ionization efficiency is defined by analogy with the
scintillation efficiency, i.e. as the ratio of the ionization yield
of nuclear recoils ($Q_{y,nr}$) to that of electron-equivalent
recoils ($Q_{y,ee}$), at the same energy:
$L_{ion}=Q_{y,nr}/Q_{y,ee}$. Combining the data of Table~1 and
Fig.~\ref{EIonYield}, one can obtain that $L_{ion}$ amounts to
0.30$\pm$0.04 at 233 keV and 0.27$\pm$0.04 at 80 keV.

Accordingly, one may conclude that the ionization yields of nuclear
recoils in liquid Ar at energies above 80 keV might be relatively
large, reaching values of the order of 10 $e^-$/keV at electric
fields exceeding 2 kV/cm. In addition, the ionization quench factor
in these conditions is about 30\%.

\section{Comparison with other experiments and with theoretical models}
\begin{figure}
\includegraphics[width=0.8\columnwidth]{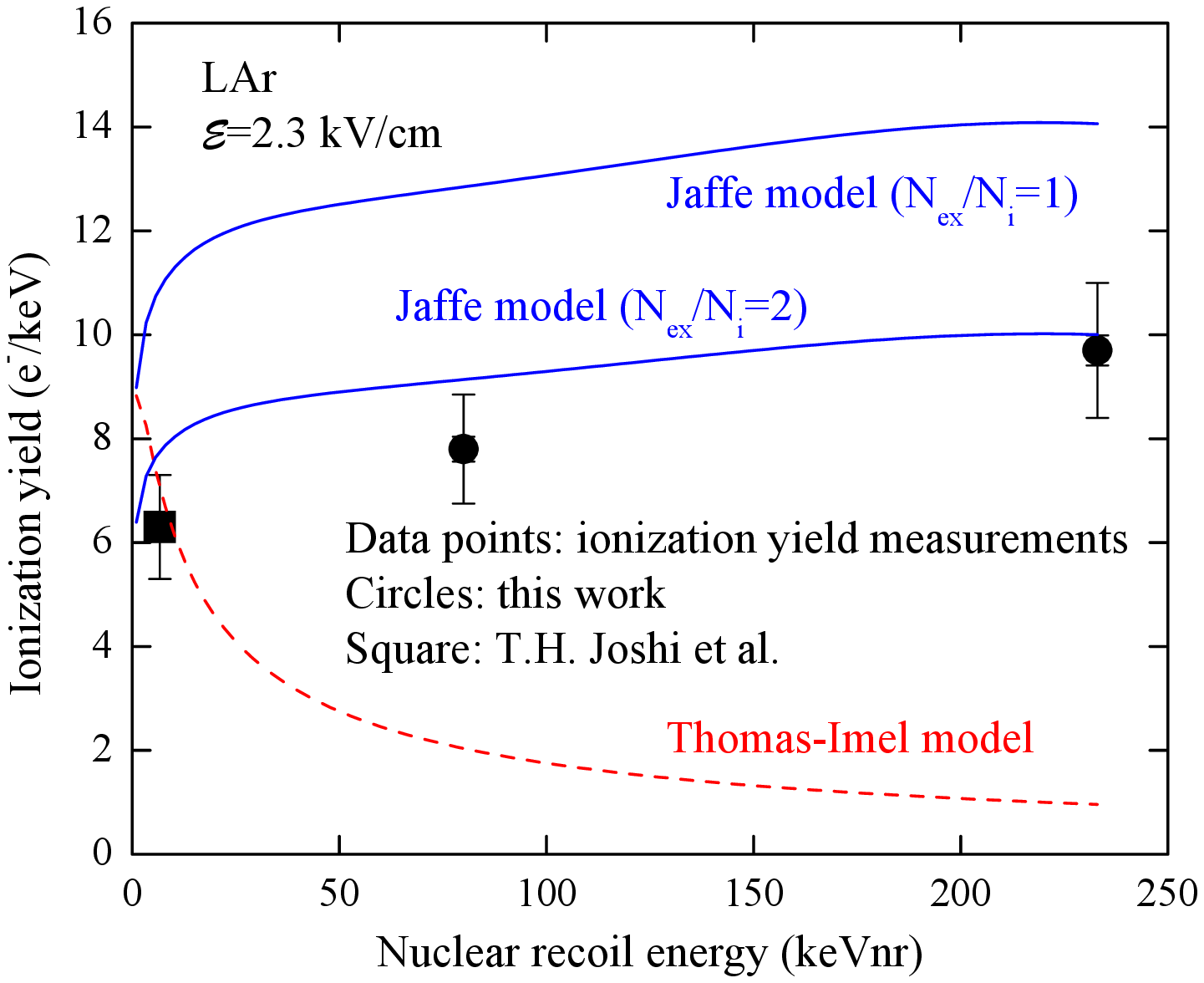}
\caption{Ionization yield of nuclear recoils in liquid Ar at 2.3
kV/cm as a function of the energy, measured in the present work
(circles) and in T.H. Joshi et al. \cite{Joshi} (square). The
theoretical predictions for the ionization yield are also shown,
obtained in the frame of Thomas-Imel (dashed curve) and Yaffe (solid
curve) models.} \label{IonYieldVsE}
\end{figure}

In Fig.~\ref{IonYieldVsE} our data on the ionization yield are
compared to those of ref. \cite{Joshi} measured at 6.7 keV. The data
point of the latter was obtained at a close field value, at 2.13
kV/cm; it was extrapolated to that of 2.3 kV/cm, using the measured
field dependence of ref. \cite{Joshi}. One can see that the
experimental data look compatible in terms of the absolute value of
the ionization yield. In addition, the ionization yield looks to be
an increasing function with energy.

On the other hand, the ionization yield measured at lower energies,
below 57 keV, in another work in the field \cite{Cao}, was reported
to be a decreasing function with energy, in contrast to our results.
The explanation of such a discrepancy might be a non-monotonic
energy dependence of the ionization yield, predicted in some
computer simulation models \cite{NEST}.

Basically there are two theoretical models that can describe the
recombination effect in liquid noble gases and consequently the
energy and field dependence of the ionization yield of nuclear
recoils: that of Thomas-Imel \cite{TIModel1,TIModel2} and that of
Jaffe \cite{JaffeModel1,JaffeModel2}.

One can see in Fig.~\ref{IonYieldVsE} that the energy dependence
measured in experiment fully contradicts to Thomas-Imel model
\cite{TIModel1,TIModel2} that predicts the decreasing function with
energy for the ionization yield. In this model,

\begin{equation}
\label{eq.3} n_e=\frac{N_i}{\xi}ln(1+\xi) ; \hspace{1ex}
\xi=\frac{N_i C}{\mathcal{E}^b} ; \hspace{1ex} N_i=\frac{f
E_0}{W_s^{min}} (1+N_{ex}/N_i)
\end{equation}

\begin{equation}
\label{eq.4} Q_y=\frac{f}{W_s^{min}} (1+N_{ex}/N_i)
\frac{ln(1+\xi)}{\xi}
\end{equation}
Here $b$ and $C$ are fitting constants, their values being taken
from ref. \cite{Joshi} for the data point at 6.7 keV;
$W_s^{min}$=19.5 eV is the minimum energy required to produce a
scintillation photon (i.e. to produce excitation or ionization) in
liquid Ar \cite{NobleRev}. The ratio of the number of excitations to
that of ionizations was taken the same as in \cite{Joshi}:
$N_{ex}/N_i$=0.2.

$f$ is the Lindhard factor (also called quench factor),
characterizing the nuclear recoil collisions; it is defined as the
energy fraction transferred to ionization ($E_i$) and excitation
($E_{ex}$):

\begin{equation}
\label{eq.5} f=(E_i+E_{ex})/E_0
\end{equation}
For electron recoils it is equal to unity. For nuclear recoils it
can be calculated in two ways: either taking the calculations of
Linhard et al. \cite{Lindhard} or using a SRIM computer program
\cite{SRIM} developed for low energy ion collisions. In the present
work we used the second method.

In contrast to Thomas-Imel model, the Jaffe model \cite{JaffeModel1}
in its compact form \cite{JaffeModel2} predicts the increasing
function with energy for the ionization yield: see
Fig.~\ref{IonYieldVsE}. In this form the model has a Birks-like
equation \cite{JaffeModel2},

\begin{equation}
\label{eq.6} n_e=\frac{N_i}{1+k_B(dE/dx)/\mathcal{E}}
\end{equation}
Here $k_B$ is a constant; it was determined from the field
dependence of the data of ref. \cite{Joshi} at 6.7 keV. The stopping
power (energy losses) for excitation and ionization, $dE/dx$, was
calculated using the SRIM program \cite{SRIM}.

The equation for the ionization yield in the frame of this Jaffe
model was derived ourselves from Eq.~6,

\begin{equation}
\label{eq.7}
Q_y=\frac{f}{[1+k_B(dE/dx)/\mathcal{E}][E_g+E_{ph}(N_{ex}/N_i)]}
\end{equation}
using Eq.~5, where $E_i=N_i E_g$ and $E_{ex}=N_{ex}E_{ph}$ (see
section 4 in ref. \cite{NobleRev} for more details of the energy
balance in liquid noble gases). Here $E_g\simeq 14.2$ eV is the band
gap in liquid Ar, which we adopt equal to that of solid Ar (see
table 2 in ref. \cite{NobleRev}); $E_{ph}$=9.7~eV is the average
energy of scintillation photon. In the lack of theoretical
and experimental data for liquid Ar, the ratio $N_{ex}/N_i$ can be
taken here either equal to that measured for nuclear recoils in
liquid Xe \cite{Sorensen11}, namely $N_{ex}/N_i$=1, or 10 times
greater than the ratio for electron recoils in liquid Ar (in the
same way as it was for Xe - see table 2 in \cite{NobleRev}), namely
$N_{ex}/N_i$=2. One can see that the Jaffe model can probably
consistently describe the experimental data in terms of the energy
dependence, and even in terms of the absolute values in the latter
case.

\section{Conclusions}

In the present work, the ionization yield of nuclear recoils in
liquid Ar has for the first time been measured at higher energies,
namely at 80 and 233 keV, using a two-phase Cryogenic Avalanche
Detector (CRAD) and DD neutron generator. The ionization yield in
liquid Ar at an electric field of 2.3 kV/cm amounted to 7.8$\pm$1.1
and 9.7$\pm$1.3 e$^-$/keV at 80 and 233 keV respectively; the
ionization quench factor amounted to 0.27$\pm$0.04 and 0.30$\pm$0.04
respectively. The Jaffe model can probably consistently describe the
energy dependence of the ionization yield, in the energy range of
6-230 keV, in contrast to that of Thomas-Imel. The results of such a
study is relevant to the energy calibration of liquid
noble gas detectors for dark matter search experiments and
understanding of the ionization yield in liquid Ar.

\acknowledgments
This work was supported in part by the Ministry of
Education and Science of Russian Federation and by the grants of the
Government of Russian Federation (11.G34.31.0047) and the Russian
Foundation for Basic Research (12-02-91509-CERN-a).

\end{document}